
\input phyzzx.tex
\pubnum={4}
\titlepage
\title{\bf Rotation Curves of Spiral Galaxies and Large Scale Structure
of Universe under Generalized Einstein Action}
\author{M. Kenmoku, E. Kitajima, Y. Okamoto}
\address{Department of Physics, Nara Women's University\break
Nara 630, Japan}
\andauthor{K. Shigemoto
\foot{E-mail address: shigemot@jpnyitp.bitnet} }
\address{Department of Physics, Tezukayama University \break
Nara 631, Japan}
\vfil
\abstract{We consider an addition of the term which is a square of the
scalar curvature to the Einstein-Hilbert action.  Under this generalized
action, we attempt to explain  i) the flat rotation curves observed in
spiral galaxies, which is usually attributed to the existence of dark
matter, and ii) the contradicting observations of uniform cosmic
microwave background and non-uniform galaxy distributions against
redshift.  For the former, we attain the flatness of velocities, although
the magnitudes remain about half of the observations.  For the latter,
we obtain a solution with oscillating Hubble parameter under uniform
mass distributions.  This solution leads to several peaks of galaxy
number counts as a function of redshift with the first peak corresponding
to the Great Wall.}
\vskip 1.0 cm
\line{PACS number(s): 04.20.-q, 98.80.Dr, 98.60.Eg \hfil }
\endpage
\def\sqr#1#2{{\vcenter{\hrule height.#2pt
    \hbox{\vrule width.#2pt height#1pt \kern#1pt
       \vrule width.#2pt}
    \hrule height.#2pt}}}
\def\square{{\mathchoice{\sqr84}{\sqr84}{\sqr{5.0}3}{\sqr{3.5}3}}}
\def\pd{\partial }
\REF\rfone{R. Utiyama and B.S. de Witt, {\it J. Math. Phys.} {\bf 3},
608 (1962).}
\REF\rftwo{H. Nariai and K. Tomita, {\it Prog. Theor. Phys.} {\bf 46},
776 (1971).}
\REF\rfthree{For example, see R. Sancisi and T.S. van  Albata, \lq\lq
IAU Symposium" No.117, p.69 (Reidel, Dordrecht, 1987).}
\REF\rffour{For example, see J.P. Ostriker, P.J.E. Peebles and A. Yahil,
{\it Astrophys. J.} {\bf 193}, L1 (1974).}
\REF\rffive{J.C.Mather {\it et al}., {\it Astrophys. J.} {\bf 354},
L37 (1990).}
\REF\rfsix{M. Geller and J. Huchra, {\it Science} {\bf 246}, 857 (1989).}
\REF\rfseven{T.J. Broadhurst {\it et al}., {\it Nature} {\bf 343},
726 (1990).}
\REF\rfeight{M. Morikawa, {\it Astrophys. J.} {\bf 369}, 20 (1991).}
\REF\rfnine{S. Weinberg, \lq\lq Gravitation and Cosmology"
(John Wiley and Sons, New York, 1972).}
\REF\rfnione{K. Stella, {\it Gen. Rel. Grav.} {\bf 9}, 353 (1978).}
\REF\rfnitwo{R. Utiyama, {\it Prog. Theor. Phys.} {\bf 72}, 83 (1984).}
\REF\rften{K.C. Freemann, {\it Astrophys. J.} {\bf 160}, 811 (1970).}
\REF\rfeleven{R.H. Sanders, {\it Astron. Astrophys.} {\bf 136}, L21
(1984).}
\REF\rftwelve{For example, see Ref.~9.}
\FIG\fgI{Rotation curves of spiral galaxies under the generalized
Einstein action. The numbers on the curves indicate the values of $a$
in Eq.~(2.23). The prediction of Newtonian theory for $a=2$ is also
given by a broken curve for comparison. }
\FIG\fgII{Scale factor of universe as a function of $\tau$ for various
values of parameters $c_0$, $\Omega_{N_0}$, and
$a_0^{\prime\prime}$~.  The parameter $c_0$ is positive for
(a) $\sim$ (d)
and negative for (e) and (f).
The values of $|c_0|$ are indicated in the figure by the letters
$a$, $b$, $c$, $d$, $e$, $f$ for $10^0$, $10^{-1}$, $10^{-2}$,
$10^{-3}$, $10^{-4}$, $10^{-5}$, respectively. }
\FIG\fgIII{(a) Scale factor of universe as a function of $\tau$
for the set of parameters in (3.31) $\sim$ (3.33)
and (b) number count of galaxies as a function of $z$ for the same
set of parameters.
The scale of the ordinate for (b) is arbitrary. }
\doublespace
\chapter{Introduction}
One of the most straightforward generalizations of the Einstein-Hilbert
action is the addition of a square terms of the scalar curvature and
Ricci tensor.  This action was introduced as counter terms to regulate
ultraviolet divergences of the Einstein theory.\refmark{\rfone} It was
also used to obtain a bounce universe, which avoids the initial
singularity of the big bang cosmology.\refmark{\rftwo}

In the present work we consider this action in order to
overcome the difficulties encountered by the standard Einstein theory
in explaining certain astrophysical and cosmological observations.  The
famous astrophysical observations that appear to be in conflict with
our expectations from the Newton's theory of gravity is the flat
rotation curves of spiral galaxies.\refmark{\rfthree}  The rotation
velocities
are constant as a function of the distance from the center of galaxy,
while a naive application of the $1/r^2$ force law implies a decline
in the velocity function.  This observation is usually accounted for
by the incorporation of dark matter.\refmark{\rffour}  In the present
work we attempt to obtain flat rotation curves from the generalized
Einstein action without relying upon dark matter.  As for the
cosmological
observations, the recent technological developments in observations
have yielded contradictory results : highly uniform cosmic microwave
background \refmark{\rffive} and non-uniform galaxy distributions in
the scales of hundreds of Mpc.\refmark{\rfsix,\rfseven}  In order
to resolve this difficulty, Morikawa introduced a non-conformal scalar
field in the action and obtained a theory with oscillating Hubble
constant.\refmark{\rfeight}  According to this theory, the mass
distribution of the universe is uniform at any moment (hence, uniform
cosmic microwave background), but the galaxiy distributions as a
function of redshift become periodic (hence, non-uniform large
scale structure of the universe) due to the oscillation in the
expansion rate of the universe.  In the present work, we follow the same
idea under the generalized Einstein action without introducing the
scalar matter.

The paper is organized as follows. In sect.~2 we discuss rotation
curves of spiral galaxies under the generalized Einstein action.
In sect.~3 we consider galaxy number counts under the generalized
Einstein action. A concluding remark is given in sect.~4.
\chapter{Rotation Curves of Spiral Galaxies under the Generalized
Einstein Action}

In this section, we consider a modification of Newtonian theory
of gravitation under the generalized Einstein action and apply this
force law to the rotation curves of spiral galaxies.  The generalized
Einstein action is given by
$$ I_g=-{1 \over 16\pi G}{\int}d^4x\sqrt{g}(R+2\Lambda+c_1 R^2
+c_2 R_{\mu \nu}R^{\mu \nu})\ , \eqn \eqI $$
where $G$ is the gravitational constant, $R$ is the scalar curvature,
$R_{\mu\nu}$ is the Ricci tensor, $\Lambda$ is the cosmological
constants, and $g$ is the negative determinant of the metric tensor.
Throughout this paper we follow the conventions of Ref.~9 and, in
particular, the speed of light is set equal to unity.  The Einstein
equation in this theory is then written as,\refmark{\rftwo}
$$R_{\mu \nu}-{1 \over 2}Rg_{\mu \nu}-\Lambda g_{\mu \nu}
+c_1 J_{\mu \nu}+c_2 K_{\mu \nu}=-8{\pi}GT_{\mu \nu}\ , \eqn \eqII $$
where $J_{\mu\nu}$ and $K_{\mu\nu}$ are defined by
$$ J_{\mu \nu}=2R(R_{\mu \nu}-{1 \over 4}Rg_{\mu \nu})
+2(R_{; \mu \nu}-g_{\mu \nu}{\square}R)\ , \eqn \eqIII $$
$$ K_{\mu \nu}=R_{; \mu \nu}-\square R_{\mu \nu}-{1 \over 2}
(\square R+R_{\alpha \beta}R^{\alpha \beta})g_{\mu \nu}
+2R^{\alpha \beta}R_{\mu \alpha \nu \beta}\ , \eqn \eqIV $$
with
$$ R_{; \mu \nu} \equiv \nabla_{\mu} \nabla_{\nu} R, \qquad
\square R \equiv g^{\alpha \beta} R_{; \alpha \beta}\ . \eqn \eqV $$

In order to obtain the modification to the Newton's $1/r^2$ law under
this theory, we consider the following static, weak field limit:
$$ g_{0 0} \cong -(1+2\phi)\ , \eqn \eqVI$$
$$ g_{i j} \cong \delta_{i j}(1+2\psi)\ , \eqn \eqVII $$
where $\phi$ and $\psi$ are functions of spatial coordinates only.
Note that $\phi$ corresponds to the gravitational potential field.
The Ricci tensor and scalar curvature are then given by
$$ R_{0 0} \cong -\bigtriangleup\phi\ , \eqn \eqVIII$$
$$ R_{i j} \cong (\partial^2_{i}\phi+\partial^2_{i}\psi
+\bigtriangleup\psi)\delta_{i j}\ , \eqn \eqIX$$
$$ R \cong 2\bigtriangleup\phi+4\bigtriangleup\psi\ . \eqn \eqX $$
Substituting these into the time-time component of Eq.~\eqII, we
obtain the equation for $\phi$ and $\psi$:
$$ \bigtriangleup[\phi+2c_1 \bigtriangleup\phi
+(4c_1 +2c_2)\bigtriangleup\psi]=4{\pi}G\rho\ , \eqn \eqXI $$
where $\rho$ is the mass density. The trace of Eq.~\eqII, on the other
hand , gives another equation for $\phi$ and $\psi$:
$$ [1+(6c_1 +2c_2)\bigtriangleup]\bigtriangleup(\phi+2\psi)
=-4{\pi}G\rho\ . \eqn \eqXII $$
We have to solve \eqXI\ and \eqXII\ simultaneously.
For the point source with
$$ \rho=M\delta^{(3)}(\vec{r})\ , \eqn \eqXIII$$
we obtain the following solution:\refmark{\rfnione, \rfnitwo}
$$ \phi=GM\left[-{1 \over r}-{1 \over 3}{e^{-m_1 r} \over  r}
+{4 \over 3}{e^{-m_2 r} \over r}\right]\ , \eqn \eqXIV$$
$$ \psi=GM\left[{1 \over r}-{1 \over 3}{e^{-m_1 r} \over r}
-{2 \over 3}{e^{-m_2 r} \over r}\right]\ , \eqn \eqXV $$
where
$$ m_1^2=-{1\over {6c_1+2c_2}}\ ,\qquad m_2^2={1\over c_2}\ .\eqn\eqXVI$$
Note that the first term in \eqXIV\ is the usual Newton potential
and that the second and third terms in \eqXIV\ correspond to its
corrections. The third term in \eqXIV\ is, however, undesirable,
since its existence does not yield the usual attractive Newton's force
law in the limit $r \rightarrow 0 $. We thus set
$$ c_2 =0\ .   \eqn\eqXVII$$
As for the constant $c_1$, a positive $c_1$ gives
$$ \phi=-GM\Big[ {1 \over r} +{1 \over 3}\Big( {{\cos \mu r} \over r}
+y{{\sin \mu r} \over r} \Big) \Big]\ , \eqn \eqXVIII $$
where $y$ is an arbitrary constant and
$$ \mu={1 \over \sqrt{6c_1}}\ ,   \eqn \eqXIX  $$
while a negative $c_1$ gives
$$ \phi=-GM\Big[ {1 \over r} +{1 \over 3} {{e^{-m_1 r}} \over r}
 \Big]\ .  \eqn \eqXX $$
Both cases give the correct $r \rightarrow 0 $ limit of the observed
gravitational constant $G_0$ with
$$ G_0={4 G \over 3}\ .        \eqn \eqXXI $$
Eq.~\eqXX, however, implies more decline of force than the $1/r^2$ law
as a function of $r$, and this is opposite to what is expected from the
rotation curves of galaxies.  We thus adopt the solution
of Eq.~\eqXVIII\ and calculate rotational velocities of typical spiral
galaxies. Taking into account the exponential mass distribution implied
by the surface
photometry data,\refmark{\rften} we approximate a spiral galaxy by a
truncated disk with mass density
$$
 \rho(\vec{r})=\cases{& $\sigma_0\ e^{-{r \over a}}\ \delta (z)\ ,\qquad
(r\le r_g)$ \cr
 & $0\ ,\qquad (r > r_g)$ \cr }\eqn \eqXXII
$$
where we take
$$ r_g=4.3 a\ [{\rm kpc}]\ , \qquad  a=2,3,4,5,6\ , \eqn \eqXXIII $$
for the truncation radius.\refmark{\rfeleven}  The value of $\sigma_{0}$
was also taken from Ref.11.  Namely, we impose the condition that the
total mass is $2.2\times10^{10} M_{\odot}$ for $a=2$, where $M_{\odot}$
is the solar mass.
The free parameter $\mu$ and $y$ in Eq.~\eqXVIII\ were chosen so that the
velocity curve becomes flat. Our choice is
$$ \mu={1 \over 50}\ \Big[ {1 \over {\rm kpc}} \Big]\ , \eqn \eqXXIV $$
$$ y=3\ .   \eqn \eqXXV $$
We thus obtain the square of the rotation velocity as a function of
the distance $r$ from the center of galaxy by
$$ v^2(r)={\sigma_{0} \over 4\pi} r \int_0^{r_g} dr^{\prime}
\int_0^{2\pi}d\theta^{\prime} r^{\prime} e^{-r^{\prime}/a}
{\pd \over \pd r}
\phi_0(|{\vec r} -{\vec r^{\prime}}|)\ , \eqn \eqXXVI $$
where $\phi_0$ is the Green's function of Eq.~\eqXVIII\ with $M=1$, and
we only consider ${\vec r}$ on the plane of the galaxy disk.

In Fig.~1, we show the results of numerical integration of Eq.~\eqXXVI\
for each value of $a$ with $r$ in the range of typical galaxy radius,
i.e., 10---100 kpc.  For a comparison, the result for $a=2$ under usual
Newtonian gravity is also shown by the broken curve in the figure.  As
is clear from
the figure, the rotation velocities become flat for $r\gsim$ 50 kpc,
while that of Newtonian gravity steadily decreases as $r$ increases.
However, the magnitudes of velocities are about half of what are
observed for each value of $a$ in Eq.~(2.23) for $r\gsim 50$ kpc.

\chapter{Large Scale Structure of Universe under the Generalized
Einstein Action}

In this section we study the cosmology under the generalized Einstein
action.
The action we consider is given by Eq.~\eqI\ with $c_2=0$.  Assuming the
homogeneous and isotropic universe, we employ the Robertson-Walker
metric:
$$ ds^2 = -dt^2+a^2(t) \Big({dr^2 \over 1-kr^2}+r^2 d{\theta}^2
+r^2 \sin^2 \theta d{\varphi}^2 \Big)\ , \eqn \eqXXVII $$
and the energy-momentum tensor $T^{\mu \nu}$ of the form
$$ T^{\mu \nu}=pg^{\mu \nu}+(p+\rho)U^{\mu}U^{\nu}\ . \eqn \eqXXVIII $$
The time-time component and space-space component of the Einstein
equation, Eq.~\eqII, are then respectively written as
$$ H^2+K-{\Lambda \over 3}-6c_1(-\dot{H}+K^2+6\dot{H}H^2-2KH^2+2H\ddot{H})
={8{\pi}G \over 3}\rho\ , \eqn \eqXXIX $$
and
$$\eqalign{ -2\dot{H}-3H^2-K+\Lambda
    &+6c_1(2\partial_t^3{H}+12H\ddot{H}+9\dot{H}^2+18\dot{H}H^2-4K\dot{H} \cr
-2KH^2-K^2)&=8{\pi}Gp\ , \cr} \eqn \eqXXX $$
where
$$ H={{\dot a} \over a}\ ,   \eqn \eqXXXI $$
and the dots stand for time derivatives.  We now introduce a few
dimensionless parameters as follows;
$$ \tau \equiv H_0 t\ ,   \eqn \eqXXXII $$
$$ \Omega_0 \equiv  {\rho_0 \over \rho_c}\ , \eqn \eqXXXIII $$
$$ k_0 \equiv { k \over H_0^2}\ , \eqn \eqXXXIV $$
$$ \lambda_0 \equiv {\Lambda \over 3 H_0^2}\ , \eqn \eqXXXV$$
where $H_0$ is the Hubble parameter at the present time,
and the
critical density $\rho_c$ is defined by
$$ \rho_c \equiv {3 \over 8\pi G} H_0^2\ .  \eqn \eqXXXVI  $$
The density parameter $\Omega_0$ have two components:
$$ \Omega_0 \equiv \Omega_{N_0} +\Omega_{R_0}\ , \eqn \eqXXXVII $$
where $\Omega_{N_0}$ and $\Omega_{R_0}$ are respectively non-relativistic
and relativistic components.  We can then write
$$ {8{\pi}G \over {3H_0^2}}\rho = \Omega_0{{\rho} \over {\rho_0}}
={\Omega_{N_0} \over a^3}+{\Omega_{R_0} \over a^4}\ , \eqn \eqXXXVIII $$
where we have set the present value of the scale factor $a_0$ to unity:
$$ a_0 \equiv a(0) =1\ .  \eqn \eqXXXIX  $$
By \eqXXXII\ $\sim$ \eqXXXV, \eqXXXVIII\  and the energy-momentum
conservation
$$ \dot{\rho}=-3\left( {\dot{a} \over a}\right)(\rho+p)\ , \eqn \eqXXXX $$
we finally reduce \eqXXIX\ and \eqXXX\ to
$$
\eqalign{
{a^{\prime\prime\prime } \over a} =
&-{1 \over 12c_0}\biggl({a \over a^{\prime }}\biggr)
\Biggl[{\Omega_{N_0} \over a^3}+{\Omega_{R_0} \over a^4}
-\biggl({a^{\prime }\over a}\biggr)^2-{k_0 \over a^2}+\lambda_0\Biggr]\cr
&-{1 \over 2}\biggl( {a \over a^{\prime }}\biggr)
\Biggl[ 2{a^{{\prime }^2} a^{\prime\prime } \over a^3}
-\biggl( {a^{\prime\prime } \over a}\biggr) ^2
-3\biggl( {a^{\prime } \over a}\biggr) ^4 \cr
& +{k_0 \over a^2}\Biggl( {k_0 \over a^2}
-2\biggl( {a^\prime \over a}\biggr) ^2 \Biggr) \Biggr]\ , \cr
} \eqn \eqXXXXI
$$
and
$$\eqalign{
{a^{\prime\prime\prime\prime} \over a} =&{1 \over 12c_0}
\Biggl[{\Omega_{R_0} \over a^4}+2{a^{\prime\prime} \over a}
+\biggl({a^{\prime} \over a}\biggr)^2+{k_0 \over a^2}
-3\lambda_0\Biggr] \cr
& -{1 \over 2}\Biggl[ 4{a^{\prime} a^{\prime\prime\prime} \over a^2}
-12{a^{{\prime}^2} a^{\prime\prime} \over a^3}
+3\biggl({a^{\prime\prime} \over a}\biggr)^2
+3\biggl({a^{\prime} \over a}\biggr)^4 \cr
&-{k_0 \over a^2}\Biggl({k_0 \over a^2}+4{a^{\prime\prime} \over a}
-2\biggl({a^{\prime} \over a}\biggr)^2\Biggr) \Biggr]\ , \cr}
\eqn \eqXXXXII $$
respectively, where
$$ c_0 \equiv c_1 H_0^2\ , \eqn \eqXXXXIII $$
and the prime stands for derivative with respect to $\tau$.
These are third-order and fourth-order differential equations for
$a(\tau)$. In particular, we consider solving Eq.~\eqXXXXII\ numerically.
For this we need {\it initial} condition of $a$,${a^{\prime }}$,
${a^{\prime \prime}}$, and
${a^{\prime \prime \prime}}$. We have chosen $a_0=1$ (Eq.~\eqXXXIX).
It then
follows that
$$     a_0^{\prime} =1\ ,  \eqn \eqXXXXIV $$
$$     a_0^{\prime \prime}=-q_0\ ,  \eqn \eqXXXXV   $$
where $q_0$ is the deacceleration parameter
$$ q_0 \equiv -{a_0^{\prime \prime} a_0 \over a_0^{\prime 2}}\ .
\eqn  \eqXXXXVI $$
By choosing values of $\Omega_0$,$k_0$,$\lambda_0$, and $q_0$ which
are observationally acceptable, we can determine
$a_0^{\prime \prime \prime}$ for each value of $c_0$ from Eq.~\eqXXXXI\
evaluated at the present time:
$$ \Omega_0-k_0+\lambda_0+6c_0\left[2a_0^{\prime\prime\prime}
-(a_0^{\prime\prime})^2+2a_0^{\prime\prime}-3
+k_0^2-2k_0\right]=1\ . \eqn \eqXXXXVII $$

In the present work, we set
$$ k_0=0\ ,     \eqn \eqXXXXVIII $$
$$ \lambda_0=0\ ,  \eqn \eqXXXXIX $$
$$ \Omega_{R_0}=10^{-5}\ ,  \eqn \eqXXXXX $$
for simplicity and use only positive $q_0$. (Non-zero values of $k_0$
and $\lambda_0$ or negative $q_0$ gave similar results unless their
values are taken to be unnaturally large.)  This leaves $\Omega_{N_0}$,
$q_0(=-a_0^{\prime \prime})$,
and $c_0$ as free parameters. In Fig.~2 we show the results of numerical
integration of Eq.~\eqXXXXII\ for several sets of values of these
parameters.
For $c_0 > 0$, we have a
de Sitter-like expansion (Fig.~2($a$)), a bounce (Curve $f$ in
Fig.~2($b$)) which is essentially the bounce solution of Ref.~2, an
expansion followed by an eventual collapse (Fig.~2($c$)), or an
expansion, shrink, and eventual de Sitter-like expansion (Curve $c$ in
Fig.~2($d$)).
For $c_0<0 $, on the other hand, we generally have
oscillation: oscillating expansion or oscillating bounce (Fig.~2($e$)
and ($f$)).  This last solution, Curve $e$ in Fig.~2($e$) in particular,
is of interest, since it might explain the observed large scale
structure of galaxy number counts as a function of red shift.
\refmark{\rfsix,\rfseven}
The number of galaxies $dN$ which are located between
the comoving distance $r$ and $r+dr$ at a fixed solid angle $d\Omega$
with absolute luminosity between $L$ and $L+dL$ is given by
$$ dN=n(L,t) a^3 r^2 dr d\Omega dL\ , \eqn \eqXXXXXI $$
where $n(L,t)$ is the number density at time $t$  with absolute
luminosity $L$, satisfying
$$ n(L,t)=\left( {a_0 \over a} \right)^3 n(L,0)\ . \eqn \eqXXXXXII $$
Since the redshift $z$ and scale factor $a$ are related by
$$ z={a_0 \over a} -1\ , \eqn \eqXXXXXIII  $$
we have
$$ r=\int_t^0 {dt \over a}=\int_t^0 {z(t^{\prime})+1 \over a_0}
dt^{\prime}\ . \eqn \eqXXXXXIV $$
By substituting \eqXXXXXII\ and \eqXXXXXIV\ into \eqXXXXXI, we have
\refmark{\rfeight}
$$ {dN \over dz d\Omega dL} \propto {\left[ \int_t^0 (z(t^{\prime})+1)
dt^{\prime} \right]^2 \over H(t)}\ . \eqn \eqXXXXXV $$
Note that Eq.~\eqXXXXXIII\ gives a differential equation for $z(t)$ :
$$ {dz \over dt} =-(1+z)H(t)\ .  \eqn \eqXXXXXVI $$
Hence, by solving Eqs.\eqXXXXII\ and \eqXXXXXVI\ simultaneously, we
obtain the number count in Eq.~\eqXXXXXV.
The results for
$$ \Omega_{N_0}=0.1\ ,   \eqn \eqXXXXXVII $$
$$ a_0^{\prime \prime}=-1.0\ ,  \eqn \eqXXXXXVIII $$
$$ c_0=-0.0002\ ,   \eqn \eqXXXXXIX $$
are shown in Fig.~3 together with the scale factor $a(\tau)$. Note
that there exist four peaks in the galaxy count for $z<0.5$
with the first one
corresponding to the location of Great Wall.\refmark{\rfsix}  Even
though the peaks are not separated with equal interval, overall
features are remarkably similar to observations.\refmark{\rfseven}

\chapter{Concluding Remark}

In this paper we discussed a possible generalization of the Einstein
theory by adding a square term, $c_1 R^2$, of scalar curvature to the
action.  Choosing appropriate values of the coefficients, we found
successful explanation of
the flat rotation curves of spiral galaxies
($c_1 H_0^2 \approx 5 \times 10^{-11}$) and the large scale structure
of universe ($ c_1 H_0^2 \approx -2 \times 10^{-4} $ ) at least
at a qualitative level.
Our theory, however, is not all faultless.  It turned out that the
coefficient $\gamma$ of the Robertson expansion
is $1/2$, and this
disagrees strongly with, for instance, the radar echo experiments
from Mercury, which implies that $\gamma$ is very close to $1$ in
accord with the Einstein theory.\refmark{\rftwelve}  Originally $R^2$
term was introduced as a result of quantum corrections,
\refmark{\rfone} so that the coefficient $c_1$ will have the
logarithmic $r$ dependence as $c_1=({\rm const.})\times \log (r/r_0)$.
This interpretation may lead to a consistent explanation of both
the rotation curves of spiral galaxies and the large scale structure
of the universe from the single effective Lagrangian with the running
coefficient $c_1$ that depends on the distance scale of interest.
The difficulty of $\gamma=1/2$ may also be avoided if $r_0$ takes
a value of the order of solar radius.

\ack
We are grateful to M. Fukugita, H. Kawai, K. Kume, and K. Uehara
for useful discussions.

\refout
\figout
\bye